\documentclass[aps,prd,psfig,amsmath,amsfonts,groupedaddress,eqsecnum]{revtex4}

\newcommand{\be}{\begin{equation}}
\newcommand{\ee}{\end{equation}}
\newcommand{\bea}{\begin{eqnarray}}
\newcommand{\eea}{\end{eqnarray}}
\newcommand{\bes}{\begin{subequations}}
\newcommand{\ees}{\end{subequations}}

\newcommand{\Nv}{\dot r}

\newcommand{\Nsa}{\left({\bf n}\cdot{\cal S}_1\right)}
\newcommand{\Nsb}{\left({\bf n}\cdot{\cal S}_2\right)}
\newcommand{\Vsa}{\left({\bf v}\cdot{\cal S}_1\right)}
\newcommand{\Vsb}{\left({\bf v}\cdot{\cal S}_2\right)}

\newcommand{\Ss}{\left({\cal S}_1\cdot{\cal S}_2 \right)}
\newcommand{\NcSa}{\left({\bf n}\times{\cal S}_1\right)}
\newcommand{\NcSb}{\left({\bf n}\times{\cal S}_2\right)}
\newcommand{\VcSa}{\left({\bf v}\times{\cal S}_1\right)}
\newcommand{\VcSb}{\left({\bf v}\times{\cal S}_2\right)}

\begin{document}



\title{Post-Newtonian gravitational radiation
and equations of motion via direct
integration of the relaxed Einstein equations. \\
IV. Radiation reaction for binary systems with spin-spin coupling}


\author{Han Wang and Clifford M. Will}
\email[]{cmw@wuphys.wustl.edu}
\homepage[]{wugrav.wustl.edu/people/CMW}
\affiliation{McDonnell Center for the Space Sciences, Department of
Physics, \\Washington University, St. Louis, Missouri 63130
}


\date{\today}

\begin{abstract}
Using post-Newtonian equations of motion for fluid bodies that include
radiation-reaction terms at 2.5 and 3.5 post-Newtonian (PN) order
($O[(v/c)^5]$ and $O[(v/c)^7]$ beyond Newtonian order), we derive the
equations of motion for binary systems with spinning bodies, including
spin-spin effects.  In
particular we determine the effects of radiation-reaction coupled to
spin-spin effects on the two-body equations of motion, and on the
evolution of the spins.  
We find that radiation damping causes a 3.5PN order,
spin-spin induced precession of the individual spins.  This contrasts with
the case of 
spin-{\it orbit} coupling, where we earlier found no effect on the spins
at 3.5PN order.
Employing the 
equations of motion and of spin precession,
we verify that the loss of total energy and total
angular
momentum induced by spin-spin effects
precisely balances the radiative flux of
those quantities calculated by
Kidder {\em et al.}  
\end{abstract}

\pacs{}

\maketitle

\section{Introduction and Summary }
\label{intro}

The relativistic effects of spin may play an important role in the
inspiral of compact binary
systems, particularly involving black holes, and may have observable
effects on the gravitational-wave signal emitted.  Spin-orbit and
spin-spin
coupling leads to precessions of the spins of the bodies and of the
orbital plane, the latter effect
resulting in modulations of the amplitude of the
gravitational waveform received at a detector.  Furthermore, spin
effects
contribute directly to the gravitational waveform, and
to the overall emission of energy and angular momentum from the
system.  

In the post-Newtonian (PN) approximation to general relativity, the effects
of spin have been derived at various levels of the PN approximation.
Formally, spin-orbit and spin-spin couplings begin to affect the equations of
motion at the first post-Newtonian order, since they behave as
${\bf S} \cdot {\bf L}/m r^4 \sim {\bf S}_1 \cdot {\bf S}_2 /m r^4 \sim
(mRV)(mrv)/mr^4 \sim (mRV)^2/mr^4 \sim (m/r^2)\epsilon$, where $m$,
$v$, $r$ and $L$ denote mass, orbital velocity, separation and orbital
anglar momentum, respectively, $R$ and $V$
denote the body's size and rotational velocity, ${\bf S}$ denotes spin or
rotational angular momentum, and $\epsilon \sim v^2
\sim V^2 \sim m/r \sim m/R$ denotes the standard small ``bookeeping parameter'' of
post-Newtonian theory (we use units in which $G=c=1$)~\cite{PNnote}.  
Spin evolution effects can also be seen to be 1PN-order
effects.  Indeed, the 1PN effects of spin have been derived by
numerous authors from a variety of points of view, ranging from formal
developments of the GR equations of motion in multipole expansions
\cite{papapetrou1,papapetrou2}, to post-Newtonian calculations
\cite{obrien}, to
treatments of linearized GR as a spin-two quantum 
theory\cite{barkerocon1,barkerocon2}.
For a review of these various approaches, see \cite{barkeroconrev}.
The effects of spin on the gravitational waveform and on the energy
and angular momentum flux were worked out by Kidder {\it et
al.}~\cite{kww,kidder}.
Post-Newtonian corrections of the leading spin terms have also been
analysed~\cite{bbf1,bbf2}.

In Paper III of this series~\cite{dire3}, we derived, from first
principles, the leading effects of spin-orbit coupling in the
equations of motion at radiation
reaction order, specifically at 3.5PN order, or $O(\epsilon^{7/2})$ beyond
Newtonian gravity.  We also showed
explicitly that radiation reaction had no effect, via spin-orbit
coupling, on the individual spins themselves.  
In this paper, we extend this analysis to spin-spin coupling.  
As before, the leading contributions occur at 3.5PN order.

We use the hydrodynamic equations of motion
derived through 3.5PN order
in Papers I and II \cite{dire1,dire2}, 
and calculate the equations of motion and
spin
evolution for two spinning, finite-sized bodies.  We restrict our attention to
contributions that involve the products of the two spins.  To this end,
for each body $A$,
we decompose velocities into a center-of-mass part and an internal
(rotational) part according to ${\bf v} = {\bf v}_A + {\bar {\bf v}}$,
and expand all gravitational
potentials about the center of mass of each body using a
similar decomposition, ${\bf x} = {\bf x}_A + {\bar {\bf x}}$, and
retain only terms that contain the product $(m {\bar x}{\bar v})_1 (m {\bar
x}{\bar v})_2 \sim S_1 S_2$.  
We do not keep terms proportional to the squares of individual spins;
these represent another class of spin effects that will be studied
elsewhere.

Adopting a specific definition of ``proper spin'' ${\cal S}_A$, as defined 
in Paper III [see
Eq. (\ref{properspin})], we find the two-body equations of motion
\begin{equation}
{\bf a} =  -\frac{m}{r^2} {\bf n} 
+ {\bf a}_{\rm PN}
+ {\bf a}_{\rm PN-SO}
+ {\bf a}_{\rm PN-SS}
+ \dots 
+ {\bf a}_{\rm 2.5PN}
+ {\bf a}_{\rm 3.5PN} 
+ {\bf a}_{\rm 3.5PN-SO}
+ {\bf a}_{\rm 3.5PN-SS}
+ \dots \,,
\label{eomsummary}
\end{equation}
where ${\bf a} = {\bf a}_1 - {\bf a}_2$ is the relative acceleration.
The 1PN spin-spin terms are standard, and are given by
\be
{\bf a}_{\rm PN-SS} = - \frac{3}{\mu r^4} \left [ {\bf n} ({\bf
{\cal S}}_1 \cdot {\bf {\cal S}}_2) + {\bf {\cal S}}_1 ({\bf n}
\cdot {\bf {\cal S}}_2 )
+ {\bf {\cal S}}_2 ({\bf n} \cdot {\bf {\cal S}}_1 )
- 5{\bf n} ({\bf n} \cdot {\bf {\cal S}}_1 ) ({\bf n}
  \cdot {\bf {\cal S}}_2 )  
\right ]  \,, 
\label{aPNSS}
\ee
and the 3.5PN spin-spin contributions, derived in this paper, are given by
\bea
{\bf a}_{\rm 3.5PN-SS}  &=& \frac{1}{r^5} \biggl \{
{\bf n} \biggl [ 
\biggl ( 287\Nv^2 - 99v^2 + \frac{541}{5}\frac{m}{r} \biggr )\Nv \Ss  
- \biggl (  2646\Nv^2 - 714 v^2 +\frac{1961}{5}\frac{m}{r} \biggr )\Nv
\Nsa \Nsb
\nonumber \\
&&
+ \biggl ( 1029\Nv^2 - 123 v^2 + \frac{629}{10}\frac{m}{r} \biggr )
\biggl(\Nsa \Vsb + \Nsb \Vsa \biggr)
- 336 \Nv \Vsa \Vsb
\biggr ]
\nonumber \\
&&
+ {\bf v} \biggl [ 
\biggl ( \frac{171}{5} v^2 - 195 \Nv^2 - 67\frac{m}{r} \biggr ) \Ss
- \biggl ( 174 v^2 -1386 \Nv^2 - \frac{1038}{5}\frac{m}{r} \biggr )
\Nsa \Nsb
\nonumber \\
&&
- 438 \Nv \biggl( \Nsa \Vsb + \Nsb \Vsa \biggr) + 96 \Vsa \Vsb 
\biggr ]
\nonumber \\
&&
+  \biggl ( \frac{27}{10} v^2 - \frac{75}{2} \Nv^2 -
\frac{509}{30}\frac{m}{r} \biggr )
\biggl(\Vsb {\bf {\cal S}}_1 + \Vsa {\bf {\cal S}}_2 \biggr)
\nonumber \\
&&
+ \biggl (  \frac{15}{2} v^2 + \frac{77}{2} \Nv^2 +
\frac{199}{10}\frac{m}{r} \biggr )
\Nv \biggl(\Nsb {\bf {\cal S}}_1 + \Nsa {\bf {\cal S}}_2 \biggr) \biggr \}
\,, 
\label{a35PNSS}
\eea
where ${\bf x} \equiv {\bf x}_1 - {\bf x}_2$, $r \equiv |{\bf x}|$,
${\bf n}
\equiv {\bf x}/r$, $m \equiv m_1 + m_2$, $\mu \equiv m_1m_2/m$,
${\bf v} \equiv {\bf v}_1 - {\bf v}_2$,
$\dot r = dr/dt = {\bf n} \cdot  {\bf v}$, 
and
${\bf {\tilde L}}_{\rm N} = {\bf x} \times {\bf v}$.
The PN, PN spin-orbit, and 2.5PN contributions are standard (see Paper
III Eqs. (1.2) and (1.5) for the formulae), the
3.5PN terms were derived in Paper II, Eq. (1.3d), 
and the 3.5PN spin-orbit terms
were derived in Paper III, Eq. (1.6).

The equations of spin evolution 
are given by
\be
{\dot {\bf {\cal S}}}_1 = 
({\dot {\bf {\cal S}}}_1)_{\rm PN-S0} 
+ ({\dot {\bf {\cal S}}}_1)_{\rm PN-SS} + 
({\dot {\bf {\cal S}}}_1)_{\rm 3.5PN-SS} \,,
\label{sdotmain}
\ee
where the PN spin-orbit terms are standard (see Paper III, Eq. (1.3)).
The PN spin-spin terms are also standard, given by
\be
({\dot {\bf {\cal S}}}_1)_{\rm PN-SS} =  -\frac{1}{r^3}
\biggl ( {\bf {\cal S}}_2 - 3\Nsb {\bf n} \biggr ) 
\times {\bf {\cal S}}_1  \,. 
\label{sdotpn}
\ee
There is no 3.5PN spin-orbit contribution (Paper III), but we find a 
3.5PN spin-spin contribution, given by
\be
({\dot {\bf {\cal S}}}_1)_{\rm 3.5PN-SS} =  \frac{\mu m}{r^5} 
\biggl ( \frac{2}{3} \Vsb + 30 \Nv \Nsb \biggr ) \NcSa \,,
\label{sdot35}
\ee
with the equations for ${\cal S}_2$ obtained by interchanging the spins.

As a check of these results, we verify explicitly that the loss of total
energy and total angular momentum (including both orbital and spin) implied
by these equations matches the energy and angular momentum radiated in
gravitational waves, as calculated by Kidder {\it et al.}\cite{kww,kidder}.

In Paper III, we found that spin-orbit contributions to radiation reaction
had no effect on the proper spin of each body, {\it i.e.} 
$({\dot {\cal S}}_1)_{\rm 3.5PN-SO} = 0$, and we argued that this made
sense, given that a spinning, axisymmetric body should not couple to
gravitational radiation.  Here, however, when the coupling between the two
spins is taken into account, there {\it is} a non-trivial
radiation-reaction effect on the spins.  Nevertheless, the effect is a pure
precession; the magnitude of the spins is unaffected.  Furthermore, if
either of the spins is aligned with the orbital angular momentum ({\it i.e.}
perpendicular to ${\bf v}$ and ${\bf n}$), the other spin is not affected by
radiation reaction.

These equations of motion do not impose any limitations on the orbits.  In
particular, they can be used to evolve the quasi-circular inspiral orbits
that are typical of those considered as sources of gravitational radiation
detectable by ground-based later-interferometric 
detectors of the LIGO-VIRGO type, as well as
highly eccentric orbits of extreme mass ratio systems
that are relevant for the proposed space-based detector, LISA.

The remainder of the paper presents details.  In Section
\ref{sec:2.5pn} we derive the equations of motion to PN order,
including spin-spin terms, and show that no spin-spin effects occur at
leading radiation-reaction, or 2.5PN order.
This section illustrates some basic features of the technique of
obtaining the spin effects from the hydrodynamical equations.
In Section \ref{sec:3.5pn} we move to 3.5PN order, where the spin-spin
radiation reaction effects first appear.  Section \ref{sec:discussion}
presents concluding remarks.

\section{Post-Newtonian and 2.5PN equations of motion and spin evolution}
\label{sec:2.5pn}

\subsection{Foundations}
\label{foundations}

As in Paper III \cite{dire3}, we
analyse a binary system consisting of balls of perfect fluid that
are sufficiently small compared to their separation that tidal interactions
(and their relativistic generalizations) can be ignored, but that are
sufficiently extended that they can support a finite rotational angular
momentum, or spin.  At Newtonian order, the result is essentially trivial:
the equation of motion for body 1 is $d^2 {\bf x}_1/dt^2 = -m_2 {\bf x}/r^3 +
O(mR^2/r^4)$, where $R$ is the characteristic size of the bodies.  Spin
plays no role whatsoever, because the Newtonian interaction does not depend
on velocity.   But at post-Newtonian order, there are velocity-dependent
accelerations of the schematic form $mv^2/r^2$, and thus, taking into
account the finite size of the body and expanding about its center of mass, 
we expect to find acceleration terms
of the form $(mVR)(mVR)^\prime/mr^4 \sim S_1 S_2/mr^4$.  
However, the combination of finite size and spin introduces an
ambiguity in the definition of the center of mass of each body.  
This has given rise to the concept of ``spin supplementary
condition'' (SSC), a statement about which center of mass definition
is being used; this concept is discussed in Paper III, Appendix A.  
It turns out that this is an issue only for spin-orbit effects; 
the choice of SSC or of center of mass 
has no effect on spin-spin effects at PN or at 3.5PN order.

We will define centers of mass and spins provisionally
using the ``conserved'', or baryonic density, given
by
\begin{equation}
\rho^* \equiv \rho \sqrt{-g} u^0 \,,
\label{rhostar}
\end{equation}
where $\rho$ is the mass energy density as measured by an observer in a
local inertial frame momentarily at rest with respect to the fluid, $g$ is
the determinant of the metric, and $u^0$ is the time component of the fluid
four-velocity.  
Assuming that $\rho$ is proportional to the baryon number density, then
conservation of baryon number leads to the
useful {\it exact} continuity equation
\begin{equation}
\partial \rho^* /\partial t + \nabla \cdot  (\rho^* {\bf v}) = 0 \,,
\label{continuity}
\end{equation}
where $v^i = u^i/u^0$ is the ordinary (coordinate) velocity of the fluid.
(Greek indices range over
spacetime values $0,\,1,\,2,\,3$, while Latin indices range over spatial
values $1,\,2,\,3$.  Henceforth, spatial vectorial quantities will be
handled using a Cartesian metric.)  
The baryonic
mass, center of mass and baryonic spin of each body in our
system are defined to  be
\bes
\begin{eqnarray}
m_A &\equiv&\int_A \rho^*\,d^3x \,,
\label{baryonmass}
\\
{\bf x}_A &\equiv& m_A^{-1} \int_A \rho^* {\bf x} \,d^3x \,,
\label{baryoncenter}
\\
{\bf S}_A &\equiv&  \int_A \rho^* {\bf {\bar x}} \times 
{\bf {\bar v}} \,d^3x \,, 
\label{baryonspin}
\end{eqnarray}
\label{baryonxs}
\ees
where ${\bf {\bar x}} = {\bf x} - {\bf x}_A$ and ${\bf {\bar v}} =
{\bf v} - {\bf v}_A$.   
We also define a two-index spin quantity
\begin{eqnarray}
S_A^{ij} &\equiv 2& \int_A \rho^* {\bar x}^{[i}{\bar v}^{j]} \,d^3x \,,
\nonumber \\
S_A^{ij} &=& \epsilon^{ijk} S_A^k \,,\qquad S_A^i = \frac{1}{2} \epsilon^{ijk}
S_A^{jk} \,,
\label{spintensor}
\end{eqnarray}
where $[\,]$ around indices denotes antisymmetrization. 
With these definitions,  the baryonic mass $m_A$ is constant, and 
the velocity, acceleration and rate of change of
spin of body $A$ are given by
\bes
\begin{eqnarray}
{\bf v}_A &=& m_A^{-1} \int_A \rho^* {\bf v} \,d^3x \,,
\label{bodyv}
\\
{\bf a}_A &=& m_A^{-1} \int_A \rho^* {\bf a} \,d^3x \,,
\label{bodya}
\\
d {\bf S}_A/dt &=&  \int_A \rho^* {\bf {\bar x}} \times {\bf {a}}
\,d^3x \,.
\label{bodysdot}
\end{eqnarray}
\label{bodyvas}
\ees
Notice that, by virtue of the definition of center of mass, the ``bar'' can
be dropped from the acceleration in Eq. (\ref{bodysdot}).

\subsection{Baryonic equations of motion and spin evolution}
\label{sec2.5pnA}

We begin by working to 1PN and 2.5PN order, reproducing the standard 
1PN formulae for spin-spin interactions, and establishing some results that
will be useful when we go on to 3.5PN order.  Since we are only interested
in radiation-reaction aspects of spin, we can ignore the 2PN terms in the
equations of motion; these produce only conservative
PN corrections to the spin equations
of motion\cite{bbf1}.  

We use the hydrodynamic equations of
motion derived in Paper II, Eqs. (2.23), (2.24a) and (2.24c), 
with all quantities expressed in terms of the
conserved density $\rho^*$.  They are given by   
\be
d^2 x^i/dt^2 = U^{,i} + a_{\rm PN}^i + a_{\rm 2.5PN}^i \,,
\label{fluid}
\ee
where $U$ is the Newtonian potential, and where
\bes
\bea
a_{\rm PN}^i &=&  v^2 U^{,i} -4v^iv^j  U^{,j}- 4 U U^{,i} - 3 v^i {\dot U}
+ 4{\dot V}^i + 8v^j V^{[i,j]} 
+ \frac{3}{2} \Phi_1^{,i} - \Phi_2^{,i}
+\frac{1}{2} {\ddot X}^{,i} \,,
\label{fluidPN}
\\
a_{\rm 2.5PN}^i &=&  
\frac{3}{5} x^j ( \stackrel{(5)}{{\cal I}^{ij}} - \frac{1}{3}
\delta ^{ij} \stackrel{(5)}{{\cal I}^{kk}} ) +
2 v^j \stackrel{(4)}{{\cal I}^{ij}}
+ 2 U^{,j} \stackrel{(3)}{{\cal I}^{ij}}
+ \frac{4}{3} U^{,i} \stackrel{(3)}{{\cal I}^{kk}}
- X^{,ijk} \stackrel{(3)}{{\cal I}^{jk}} 
\,,
\label{fluid2.5PN}
\eea
\label{fluidterms}
\ees
where commas denote partial derivatives, overdots denote partial time
derivatives, and $(n)$ above quantities denotes
the number of total time derivatives.  
The potentials used here and elsewhere in the paper are given by
the general definitions
\bea
\Sigma (f) &\equiv& \int \frac{\rho^*(t,{\bf x}^\prime) 
f(t,{\bf x}^\prime)}{|{\bf x} - {\bf x}^\prime |}
d^3x^\prime \,,
\nonumber \\
X(f) &\equiv& \int \rho^*(t,{\bf x}^\prime) f(t,{\bf x}^\prime)
|{\bf x} - {\bf x}^\prime | d^3x^\prime \,,
\label{genpotentials}
\eea
with specific potentials given by 
\bea
U &=& \Sigma(1) \,, \quad
V^i = \Sigma(v^i) \,,
\nonumber \\
\Phi_1 &=& \Sigma(v^2) \,, \quad
\Phi_2 = \Sigma(U) \,,
\nonumber \\
X &=& X(1) \,,  \quad X^i  = X(v^i) \,. 
\label{potentials}
\eea
The multipole moment of the system ${\cal I}^{ij}$, 
as well as additional moments, 
${\cal J}^{ij}$, ${\cal J}^{ijk}$, and ${\cal M}^{ijkl}$, that
will be relevant at 3.5PN order, are defined in Paper III,
Appendix C (see also Eq. (\ref{moments})); note that there are no explicit
spin-spin terms in ${\cal I}^{ij}$, to the
PN order considered.

We now multiply the equation of motion (\ref{fluid}) by $\rho^*$ and 
integrate over body 1, expressing the variables ${\bf x}$ and ${\bf v}$ as
${\bf x}={\bf x}_A + {\bf {\bar x}}$
and
${\bf v}={\bf v}_A + {\bf {\bar v}}$, 
where $A = 1,\,2$, depending on the body in which the point lies.   
To get the acceleration of body 1, we divide the result by
$m_1$.  We use
Eqs. (\ref{baryonxs}) and (\ref{bodyvas}) to simplify where possible.  
We expand the various
potentials in powers of ${\bar x}/r$, and keep only 
terms proportional to the product of ${\bar v} \times {\bar x}$ for 
one body with ${\bar v} \times {\bar x}$ for the other body.

In Paper III, we also kept internal terms proportional to ${\bar v}^2$
and used virial relations
derived in Paper III, Appendix E to simplify expressions dependent on the
internal structure of each body.  
While the use of those virial theorems generated spin-orbit terms at PN
order, it turns out that they generate no spin-spin terms at
this order.  We will deal with the effect of virial theorems on 3.5PN
spin-spin terms in Sec. \ref{sec:3.5pn}.

The Newtonian term gives $a_{\rm N}^i = -m_2 x_{12}^i/r^3$, where, in this
paragraph, to avoid confusion, we denote $x_{12}^i \equiv x_1^i - x_2^i$, 
$r \equiv |{\bf x}_{12}|$, and ${\bf n} = {\bf x}_{12}/r$.  
The only PN terms in Eq. (\ref{fluidPN}) that can
have a $v$ in one body and a $v$
in the other body and that could therefore
lead to a spin-spin effect are the terms
$-3 v^i {\dot U}$ and $8v^j V^{[i,j]}$.  Keeping only the relevant terms,
the first of these gives, for example,
\bea
-3\frac{1}{m_1} \int_1 \rho^* v^i {\dot U} d^3x &=& -3\frac{1}{m_1} 
 \int_1 \rho^* (v_1^i + {\bar v}^i)
  d^3x
  \nonumber \\
  && \times
  \int_2 \rho^{*\prime} (v_2^j + {\bar v}^{\prime j}) 
  \biggl [ \frac{n^j}{r^2} + ({\bar x}-{\bar x}^\prime)^k \nabla^k \left (
  \frac{x_{12}^j}{r^3} \right )
  + \frac{1}{2} ({\bar x}-{\bar x}^\prime)^{k} ({\bar x}-{\bar
  x}^\prime)^{l}\nabla^{k}\nabla^{l} \left (
    \frac{x_{12}^j}{r^3} \right ) + \dots \biggr ]
  d^3x^\prime 
  \nonumber \\
  && 
  = \frac{3}{4m_1} S_1^{ki} S_2^{lj} \nabla^{k}\nabla^{l} \left (
      \frac{x_{12}^j}{r^3} \right ) \,,
\eea
where unprimed (primed) barred variables are in body 1(2).  We also 
assume that each body is in stationary equilibrium, with 
${\dot I}_A^{ij}=(d/dt)\int_A \rho^* {\bar x}^i{\bar x}^j d^3x=0$, so that
$\int_A \rho^* {\bar x}^i {\bar v}^j d^3x =  S_A^{ij}/2$.

In the combination of PN terms $4{\dot V}^i + \frac{1}{2} {\ddot X}^{,i}$
in Eq. (\ref{fluidPN}), the time derivatives generate accelerations inside
the potentials.
To the order needed for our purposes, we must therefore
substitute the Newtonian and
2.5PN continuum 
terms for those accelerations and carry out the same procedures for the
integrals as described
above.  However, while this produces spin-orbit effects (Paper III), 
it produces {\bf no} spin-spin terms.

The resulting 1PN spin-spin contribution to the equation of motion for body 1 is
\be
(a_1^i)_{\rm PN-SS} =
\frac{3}{m_1 r^4} \left ( S_1^{ij} S_2^{kj} n^k + S_2^{ij} S_1^{kj} n^k
+ S_1^{kj} S_2^{kj} n^i - 5 S_1^{kj} S_2^{lj} n^in^kn^l \right ) 
\,.
\label{a1SS}
\ee
Substituting Eq. (\ref{spintensor}) 
and calculating the relative acceleration gives
\be
{\bf a}_{\rm PN-SS} = - \frac{3}{\mu r^4} \left [ {\bf n} ({\bf
{S}}_1 \cdot {\bf {S}}_2) + {\bf {S}}_1 ({\bf n}
\cdot {\bf {S}}_2 )
+ {\bf {S}}_2 ({\bf n} \cdot {\bf {S}}_1 )
- 5{\bf n} ({\bf n} \cdot {\bf {S}}_1 ) ({\bf n}
  \cdot {\bf {S}}_2 )  
\right ]  \,, 
\label{aPNSSbaryon}
\ee
where the spins here are baryonic spins.    

Turning to the 2.5PN terms, Eq. (\ref{fluid2.5PN}),  
the integrations lead to no explicit 
spin-spin terms, so that  
the 2.5PN relative acceleration
terms are given by
\be
(a^i)_{\rm 2.5PN} =
\frac{3}{5} x^j ( \stackrel{(5)}{{\cal I}^{ij}} - \frac{1}{3}
\delta ^{ij} \stackrel{(5)}{{\cal I}^{kk}} ) +
2 v^j \stackrel{(4)}{{\cal I}^{ij}}
- \frac{1}{3} \frac{m}{r^2} n^i \stackrel{(3)}{{\cal I}^{kk}}
- 3 \frac{m}{r^2} n^in^jn^k \stackrel{(3)}{{\cal I}^{jk}}
\,.
\label{a2.5pn}
\ee
However, when we work at 3.5PN order, 
even though the multipole moments themselves
contain no explicit spin-spin terms, 
time derivatives acting on them {\bf will} produce spin-spin contributions
via the PN spin-spin terms in the equations of motion.

We calculate the evolution of the spin in a similar manner.  Starting with
$dS_1^i/dt = \epsilon^{ijk} \int_1 \rho^*{\bar x}^j a^k d^3x$, 
we expand about the
baryonic centers of mass, keeping only terms that depend on a product
of ${\bar x} \times {\bar v}$ for each body.  
At 1PN order, the only term in Eq.
(\ref{fluidPN}) that
contributes is $8v^j V^{[i,j]}$. The result, at
1PN order is
\be
({\dot {\bf {S}}}_1)_{\rm PN-SS} =  -\frac{1}{r^3}
\biggl ( {\bf {S}}_2 - 3({\bf n}\cdot {\bf S}_2) {\bf n} \biggr )
\times {\bf {S}}_1  \,,
\label{sdotpnbaryon}
\ee
where again these are baryonic spins.
At 2.5PN order, there is no spin-spin contribution.

\subsection{The proper spin}
\label{sec:properspin}

In Paper III, we defined the
proper spin of each body to be
\bea
{\cal S}_1^i &\equiv&
  S_1^i \biggl (1 + \frac{1}{2} v_1^2 + 3 \frac{m_2}{r} \biggr )
  -\frac{1}{2} [{\bf v}_1 \times ( {\bf v}_1 \times {\bf S}_1)]^i 
   - S_1^i \stackrel{(3)}{{\cal I}^{jj}} + S_1^j \stackrel{(3)}{{\cal
  I}^{ij}} \,,
  \nonumber \\
{\bf {\cal S}}_2 &\equiv& (1 \rightleftharpoons 2 ) \,.
\label{properspin}
\eea

The post-Newtonian corrections in Eq. (\ref{properspin}) arise from
transforming the baryonic 
spin from our coordinate frame to a suitable inertial frame
comoving with the spinning body.  The 2.5PN terms 
involving time derivatives of ${{\cal I}^{ij}}$
arise from the fact that the equations of motion at 2.5PN order may be
written in various gauges, such as Burke-Thorne gauge~\cite{MTW} 
(in which the 2.5PN radiation
reaction terms in the acceleration are given by $a^i_{\rm 2.5PN} =
\frac{2}{5} x^j d^5 {\cal I}^{<ij>}/dt^5$), or Damour-Deruelle
gauge~\cite{DD81,Damour82} (the gauge used in this paper). 
Including the 2.5PN terms as in (\ref{properspin}) is equivalent to
defining our spins in the Burke-Thorne gauge.  In any case, such quantities as
angular momentum and energy are well-defined only up to the order at
which they conserved, and one is free to add 2.5PN and 3.5PN
order terms to them without affecting their fundamental conserved properties;
including the 2.5PN terms in 
(\ref{properspin}) has the property that radiation reaction effects in the
proper spin do not appear (if they appear at all) until 3.5PN order.
With this definition, the proper spins ${\bf {\cal S}}_A$ also satisfy the
standard spin-orbit precession equations, Paper III, Eq. (1.3).
Transforming from our baryonic spin to the proper spin will generate some
spin-spin terms at 3.5PN order.

\subsection{Conserved total energy and angular momentum}
\label{centermass}

The Newtonian and PN spin-spin terms in the equation of motion
(\ref{aPNSSbaryon}), and the PN spin-spin terms in the spin precession equation
(\ref{sdotpnbaryon}),
together imply conservation of the total energy and angular momentum of the
system, given to Newtonian and PN spin-spin order by
\bes
\bea
E &=& \mu \left ( \frac{1}{2}v^2 - \frac{m}{r}
\right ) - \frac{1}{r^3} \left [\Ss - 3\Nsa\Nsb \right ]  
\,,
\\
{\bf J} &=&
\mu {\bf {\tilde L}}_{\rm N} 
+ {\bf {\cal S}} \,,
\eea
\label{EJrelative}
\ees
where ${\cal S} = {\cal S}_1 + {\cal S}_2$.
Notice that there is no PN
spin-spin contribution to the total angular momentum.  In Eq.
(\ref{EJrelative}) we have converted the baryonic spins to the proper
spins; the PN corrections in Eq.
(\ref{properspin}) do not introduce new spin-spin contributions to
$E$ or $J$ to the required order, and the 2.5PN corrections can always
be dropped as meaningless terms that have no effect on the conserved
quantities.

In Sec.  \ref{compareflux} we will use these expressions
together with the 3.5PN equations of motion to compare $\dot E$ and $\dot
{\bf J}$ with the corresponding fluxes of radiation to infinity.

\section{3.5PN equations of motion and spin evolution}
\label{sec:3.5pn}

\subsection{Equation of motion}

To obtain the 3.5PN contributions to the equations of motion including 
spin-spin coupling
terms, we take the 3.5PN fluid expressions shown in Eq. (D4) of Paper
III,
multiply by $\rho^*$, and integrate over body 1.  We follow
the same procedure as in Sec. \ref{sec2.5pnA}, expanding potentials about the
baryonic centers of mass of the bodies, keeping 
only spin-spin terms (terms involving products of ${\bar x}\times{\bar v}$ 
for one body with that for the other
body). 
Most terms in Eq. (D4) of Paper III 
make only point-mass or spin-orbit contributions; the only terms that
can possibly
produce non-trivial spin-spin terms in either the acceleration or the spin
evolution are:
\bea
\delta a^i_{\rm 3.5PN} &=& 
-8v^kV^{k,j} \stackrel{(3)}{{\cal I}^{ij}}
+\frac{16}{3}v^jV^{[i,j]}\stackrel{(3)}{{\cal I}^{kk}}
+8(v^jV^{k,i}-v^lX^{[i,l]jk})\stackrel{(3)}{{\cal I}^{jk}}
-\frac{16}{45} x^jv^k \epsilon^{qjk} \stackrel{(5)}{{\cal J}^{qi}}
\nonumber \\
&&
+\frac{2}{45} \left (2({\bf v} \cdot {\bf x}) \epsilon^{qik} 
-2x^iv^j \epsilon^{qjk}
        + 5x^jv^i \epsilon^{qjk} + 12x^jv^k \epsilon^{qij}
        +4x^kv^j \epsilon^{qij} \right )\stackrel{(5)}{{\cal J}^{qk}}
\nonumber \\
&&
+\frac{2}{9} (4v^jv^k \epsilon^{qij} - v^2 \epsilon^{qik} )
        \stackrel{(4)}{{\cal J}^{qk}}
-\frac{2}{9} (v^iU^{,j}+2v^jU^{,i}+V^{j,i}+{\dot X}^{,ij} )
        \epsilon^{qjk}\stackrel{(3)}{{\cal J}^{qk}}
\nonumber \\
&&
+\frac{1}{15} v^j (\epsilon^{qjk}\stackrel{(5)}{{\cal J}^{qik}}
-\epsilon^{qik} \stackrel{(5)}{{\cal J}^{qjk}}
-\epsilon^{qij}\stackrel{(5)}{{\cal J}^{qkk}} )
-\frac{1}{6} v^i \stackrel{(4)\quad}{{\cal M}^{kkjj}}
+\frac{2}{3} v^j \stackrel{(4)\quad}{{\cal M}^{ijkk}}
\,,
\label{3.5PNpart}
\eea
where, to the required order,
\bea
{\cal I}^{ij} &=& \mu x_{12}^i x_{12}^j \,,
\nonumber \\
{\cal J}^{ij} &=& -\eta \delta m {\tilde L}_{\rm N}^i x_{12}^j
-\frac{1}{2} \eta (3 \Delta^i x_{12}^j - \delta^{ij} {\bf \Delta}\cdot {\bf
x_{12}} ) \,,
\nonumber \\
{\cal J}^{ijk} &=& \eta m (1-3\eta) {\tilde L}_{\rm N}^i x_{12}^{j}x_{12}^{k}
+ \eta (2\xi^i x_{12}^{j}x_{12}^{k} 
-  {\bf \xi} \cdot {\bf x_{12}} \delta^{i(j} x_{12}^{k)}) \,,
\nonumber \\
{\cal M}^{ijkk} &=& \eta m (1-3\eta)r^2 \left ( v_{12}^{i}v_{12}^{j} -
\frac{m}{3r}n_{12}^{i}n_{12}^{j} \right ) - \frac{1}{6} \eta m^2 r (
n_{12}^{i}n_{12}^{j} - 3\delta^{ij} ) - 2\eta ({\bf x}_{12} \times {\bf
\xi})^{(i} v_{12}^{j)}
\,,
\label{moments}
\eea
where
$\delta m = m_1-m_2$, $\eta=\mu/m$,
$\xi^i = (m_2/m_1)S_1^i + (m_1/m_2)S_2^i$, 
and $\Delta^i = m(S_2^i/m_2-S_1^i/m_1)$.
Spin-spin contributions come from terms such as $v^j V^{k,i}d^3 {\cal
I}^{jk}/dt^3 $, with a
${\bar v}$ in body 1 and a ${\bar v}$ in body 2 together with suitable
${\bar x}$'s.  They also come from terms involving the current moments 
${\cal J}^{qk}$, where a single spin generated by the prefactor (eg.
$x^i v^j \epsilon^{qjk}$, or $v^i U^{,j}\epsilon^{qjk}$) is multiplied
by the spin of the other body that appears in ${\cal J}^{qk}$.  
The terms involving $d^4{\cal J}^{qk}/dt^4$ 
$d^5{\cal J}^{qpk}/dt^5$ and $d^4{\cal M}^{pqkk}/dt^4$ don't 
generate spin-spin terms
in the equation of motion (no free $\bar x$ to go with a velocity), but {\it
do} generate terms in the spin evolution.    

In addition, when the
prefactor of $d^3{\cal J}^{qk}/dt^3$ is integrated over body 1, it yields a
prefactor given by $-(2/9)(4{\cal H}_1^{(ij)} - 3{\cal K}_1^{ij})$, where 
\be
{\cal H}_A^{ij} \equiv \int_A \int_A \rho^* {\rho^*}^\prime
\frac{{\bar v}^{\prime i} (x-x^\prime)^j}{|{\bf x} - {\bf x}^\prime |^3} d^3x
d^3x^\prime \,, \quad
{\cal K}_A^{ij} \equiv \int_A \int_A \rho^* {\rho^*}^\prime
\frac{{\bf {\bar v}}^\prime \cdot ({\bf x} - {\bf
x}^\prime)(x-x^\prime)^i(x-x^\prime)^j}{|{\bf x} - {\bf x}^\prime |^5} d^3x
d^3x^\prime \,.
\ee
However, a virial theorem derived from the requirement that 
${\dddot I}_A^{ij}=0$ gives, to the required PN order, 
\be
2{\cal H}_1^{(ij)} - \frac{3}{2}{\cal K}_1^{ij} = -\frac{3}{2}
\frac{m_2}{r^3} S_1^{k(i}n^{j)k} \,.
\ee
(See Paper III, Appendix E, for a discussion of virial relations.)
This spin term, combined with those generated directly by the potentials,
and multiplied by the appropriate spin term in $d^3{\cal J}^{qk}/dt^3$,
gives a 3.5PN spin-spin term.  This is the only place where the virial
theorems play a role in spin-spin effects.

In addition, the combination of 1PN terms $ 4{\dot V}^i
+\frac{1}{2} {\ddot X}^{,i}$ in Eq. (\ref{fluidPN}),
will generate accelerations
whose 2.5PN terms will produce 3.5PN contributions; however these
produce no spin-spin terms.
We must also re-express the 1PN spin-spin terms of Eq. (\ref{a1SS}) in terms
of the proper spin of Eq. (\ref{properspin}); the 2.5PN contributions there
will generate 3.5PN spin-spin
terms in the equation of motion.  Finally, in the 2.5PN
accelerations of Eq. (\ref{a2.5pn}), we must include the 1PN 
spin-spin terms in the equations of motion
that are generated by the many time derivatives; the explicit 1PN
corrections to those moments do not contain spin-spin terms. 

The result for the 3.5PN acceleration of body 1 is an expression
too lengthy to reproduce here.  
Calculating $a_1^i - a_2^i$ and converting 
all variables to relative coordinates using 
${\bf x}_1 = (m_2/m){\bf x}$ and ${\bf x}_2=-(m_1/m){\bf x}$, we
obtain Eq. (\ref{a35PNSS}).

\subsection{Spin evolution}

We now want to calculate the evolution of the proper spin ${\bf {\cal S}}_1$
to 3.5PN order.  A time derivative of Eq. (\ref{properspin}) gives
\bea
{\dot {\cal S}}_1^i &=& {\dot S}_1^i 
\biggl (1 + v_1^2 + 3 \frac{m_2}{r} \biggr )
   -\frac{1}{2} v_1^i ( {\bf v}_1 \cdot {\dot {\bf S}}_1)
     \nonumber\\
&&
+ S_1^i \left ( 2 {\bf v}_1 \cdot {\bf a}_1 -3\frac{m_2 {\dot r}}{r^2} \right )
-\frac{1}{2} a_1^i ({\bf v}_1 \cdot {\bf S}_1)
-\frac{1}{2} v_1^i ({\bf a}_1 \cdot {\bf S}_1)
\nonumber \\
       && 
 - S_1^i \stackrel{(4)}{{\cal I}^{jj}} 
 + S_1^j \stackrel{(4)}{{\cal I}^{ij}}
 - {\dot S}_1^i \stackrel{(3)}{{\cal I}^{jj}} 
 + {\dot S}_1^j \stackrel{(3)}{{\cal I}^{ij}} \,.
\label{dotproperspin}
\eea
We repeat the method of Sec. \ref{sec2.5pnA} to determine the contributions
of 3.5PN fluid terms to the time derivative of the baryonic spin ${\bf S}_1$,  
by calculating $\epsilon^{ijk}\int_1 \rho^* {\bar x}^j a_{\rm 3.5PN}^k d^3x$.
Only the terms displayed in Eq. (\ref{3.5PNpart}) 
will contribute spin-spin terms.   
Notice that, as we have discussed,
the 2.5PN contribution to ${\dot S}_1^i$ (actually a spin-orbit term)
cancels the first two terms in the
last line of Eq. (\ref{dotproperspin}).  For ${\bf a}_1$, which appears in
the 1PN terms in Eq. (\ref{dotproperspin}), 
we must substitute the 2.5PN equations of motion; however these make
no spin-spin contribution.   For ${\dot
S}_1^i$ in the final 2.5PN terms in Eq. (\ref{dotproperspin}) we must
substitute the 1PN spin-spin precession equations; finally we must use Eq.
(\ref{properspin}) to convert from ${\bf S}_1$ in the 1PN spin-spin
terms back to the proper spin ${\bf {\cal S}}_1$; the 2.5PN terms
there will generate 3.5PN contributions to the spin evolution.

The result is the 1PN spin precession of Eq. (\ref{sdotpn}), plus a
lengthy 3.5PN expression.
However, using the fact that, to lowest order ${\dot {\bf {\cal S}}}_1 = 0$,
together with the identities listed in Appendix \ref{app:timederiv}, it is
straightforward to show that our lengthy 3.5PN expression is almost,
but not quite, a pure
total time derivative, given by
\bea
({\dot {\bf {\cal S}}}_1)_{\rm 3.5PN-SS} &=&  \frac{\mu m}{r^5}
\biggl [ \frac{2}{3} \Vsb + 30 \Nv \Nsb \biggr ] \NcSa 
\nonumber \\
&& + \frac{d}{dt} \biggl [ \frac{\mu}{15r^3} \biggl \{
 ({\bf {\cal S}}_1 \times {\bf {\cal S}}_2)
\left [ (14-\alpha)v^2 -(42-3\alpha){\dot r}^2 + (21+\alpha)\frac{m}{r} \right ]
\nonumber \\
&&
- 10 {\bf n} [{\bf n}\cdot ({\bf {\cal S}}_1 \times
{\bf {\cal S}}_2)] \left [ 3(1+\alpha)v^2 -15(1+\alpha){\dot r}^2 
- (2-\alpha)\frac{m}{r} \right ]
\nonumber \\
&&
-90\alpha {\dot r}{\bf n} [{\bf v}\cdot ({\bf {\cal S}}_1 \times
{\bf {\cal S}}_2)]
-45(1+2\alpha) {\dot r} {\bf v}[{\bf n}\cdot ({\bf {\cal S}}_1 \times
{\bf {\cal S}}_2)]
-5(1-8\alpha) {\bf v}[{\bf v}\cdot ({\bf {\cal S}}_1 \times
{\bf {\cal S}}_2)]
\nonumber \\
&&
+3 \Nsb \NcSa \left [ 3(1+\alpha)v^2 - 15(1+\alpha){\dot r}^2 + 
(56+\alpha)\frac{m}{r} \right ]
+27 (1+\alpha) {\dot r} \Vsb \NcSa 
\nonumber \\
&&
- 9(2-3\alpha){\dot r} \Nsb \VcSa
+3(1-4\alpha) \Vsb \VcSa \biggr \} \biggr ]
\,,
\label{spindotbig}
\eea
where $\alpha = m_1/m_2$.
The total time derivative can be eliminated by
moving it to the left-hand side and absorbing it into a 
redefined proper spin ${\bf {\cal S}}_1$, which differs from the original by
meaningless 3.5PN correction terms.  This is the same philosophy by
which we absorbed the 2.5PN terms into the 
initial definition of proper spin in Eq. (\ref{properspin}).  
In the spin-orbit case of Paper
III, {\bf all} the 3.5PN terms could be so absorbed.  However, in the
spin-spin case, we find that this is not the case, and there is a
residual contribution to the spin evolution, given by the first line
in Eq. (\ref{spindotbig}).  Because there is no
unique way to absorb total time derivatives into ${\dot {\bf {\cal
S}}}_1$, the expression for this residual is not unique, although its
time average over an orbit is.  Notice that the residual term is 
orthogonal to
the spin.  In other words, the magnitude of the proper
spin of body 1 is not
affected by radiation reaction to 3.5PN order, again not surprising
for a spinning axisymmetric body.  

Redefining the proper spin, we obtain finally Eq. (\ref{sdot35}).
This represents a pure precession of the spin of body 1 about the
radial direction ${\bf n}$; however, if the companion spin is
perpendicular to the orbital plane, there is no effect.  

\subsection{Comparison with fluxes of energy and angular momentum}
\label{compareflux}

The fluxes of energy and angular momentum in gravitational waves from a binary
with spin-spin interactions were derived by Kidder {\em et al.}
\cite{kww,kidder}, and are given by
\bea
\frac{dE}{dt} &=& {\dot E}_{\rm N}  + {\dot E}_{\rm SS} \,,
\nonumber \\
\frac{d{\bf J}}{dt} &=& {\dot {\bf J}}_{\rm N} + {\dot {\bf J}}_{\rm SS} \,,
\eea
where we include only the lowest-order ``Newtonian'' and 1PN spin-spin
contributions, given by
\bes
\bea
{\dot E}_N &=& -\frac{8}{15} \frac{\eta^2 m^4}{r^4} (12v^2 -11{\dot r}^2) \,,
\label{ENflux}
\\
{\dot E}_{SS} &=&  
-\frac{4}{15} \frac{\eta m^2}{r^6} \biggl [ 
\Ss(141v^2-165\Nv^2)
+\Nsa\Nsb(807\Nv^2-504v^2)
\nonumber
\\
&&
+71\Vsa\Vsb-171\Nv\Vsa\Nsb-171\Nv\Vsb\Nsa \biggl ] \,,
\label{ESSflux}
\\
{\dot {\bf J}}_N &=& -\frac{8}{5} \frac{\eta^2 m^3}{r^3} 
{\bf {\tilde L}}_{\rm N} \left ( 2v^2 - 3{\dot r}^2 + 2\frac{m}{r} \right )
\,,
\label{JNflux}
\\
{\dot {\bf J}}_{SS} &=& 
\frac{2}{5}
\frac{\mu}{r^4} \biggl\{
\NcSa \frac{m}{r}\biggl[6\Nsb\Nv - 5\Vsb\biggl]
\nonumber
\\
&&
+\NcSb\frac{m}{r}\biggl[6\Nsa\Nv - 5\Vsa\biggl]
\nonumber
\\
&&
+\VcSa \biggl[ \Nsb \left ( 18v^2 -30\Nv^2 + 11\frac{m}{r}\right )
+6\Vsb\Nv \biggl ]
\nonumber
\\
&&
+\VcSb \biggl[ \Nsa \left ( 18v^2 -30\Nv^2 + 11\frac{m}{r}\right )
+6\Vsa\Nv \biggl ]
\nonumber
\\
&&
+ \frac{1}{r} {\bf {\tilde L}}_{\rm N} 
\biggl [ \Ss ( 60\Nv^2 -24v^2-46\frac{m}{r} )
+5\Nsa\Nsb (24 v^2-84\Nv^2+36\frac{m}{r})
\nonumber
\\
&&
+90\Nv\left (\Nsa\Vsb+\Vsa\Nsb \right )
-12 \Vsa\Vsb \biggr ]
\biggl\}
\,.
\label{JSSflux}
\eea
\label{EJfluxes}
\ees
Kidder \cite{kidder} did not determine ${\dot {\bf J}}_{SS}$
explicitly, but rather left it in the form of 
\be
{\dot J}_{SS}^i = -\frac{2}{5} \mu \epsilon^{ijk} \left \{ 
\stackrel{(2)\quad}{{\cal I}^{<jl>}}({\bf a}_{\rm N})
\stackrel{(3)\quad}{{\cal I}^{<kl>}}({\bf a}_{\rm PN-SS})
+\stackrel{(2)\quad}{{\cal I}^{<jl>}}({\bf a}_{\rm PN-SS})
\stackrel{(3)\quad}{{\cal I}^{<kl>}}({\bf a}_{\rm N})
+ \frac{16}{9} \stackrel{(2)\quad}{{\cal J}_{\rm S}^{<jl>}}({\bf a}_{\rm N})
\stackrel{(3)\quad}{{\cal J}_{\rm S}^{<kl>}}({\bf a}_{\rm N})
\right \} \,,
\ee
where the mass and current multipole moments are given by Eqs.
(\ref{moments}), where 
the angular brackets around indices denote the symmetric,
trace-free part.  The notation $({\bf a}_{\rm N})$ or 
$({\bf a}_{\rm PN-SS})$ denotes which acceleration, 
Newtonian or spin-spin, is to be used for the acceleration generated by the
time derivatives, and the subscript ${\rm S}$ denotes the 
spin part of the current moment ${\cal J}$.
Following this procedure and keeping
only terms involving the product of $S_1$ with $S_2$, we obtain Eq.
(\ref{JSSflux}).

We now calculate the time derivative of the energy and angular momentum
expressions (\ref{EJrelative}), and substitute the 
equations of motion, PN spin-spin, 2.5PN point-mass and 3.5PN spin-spin 
terms, along with the 1PN and 3.5PN spin precession equations.  
After recovering the fact that all 1PN 
spin-spin contributions cancel,
leaving $E$ and ${\bf J}$ conserved to that order, we find that the changes
in $E$ and ${\bf J}$ due to 2.5PN and 3.5PN spin-spin radiation reaction 
are obtained from the following expressions,
\bea
{\dot E} &=& \mu {\bf v} \cdot ({\bf a}_{\rm 2.5PN} + {\bf a}_{\rm
3.5PN-SS}) \,,
\nonumber \\
{\dot {\bf J}}  &=& \mu {\bf x} \times ({\bf a}_{\rm 2.5PN} + {\bf a}_{\rm
3.5PN-SS}) + {\dot {\bf {\cal S}}}\,.
\label{dotEJ}
\eea
Initially, the results do not
match the flux expressions above.  However, by making use of the identities
listed in Appendix \ref{app:timederiv}, we can show that the difference between
the expressions in all cases is a total time derivative.  These can thus be
absorbed into meaningless 2.5PN and 3.5PN corrections to the definition of
total energy and angular momentum.  Notice that the residual 3.5PN precession
term from ${\dot {\bf {\cal S}}}$ given by the sum of 
$({\dot {\bf {\cal S}}}_1)_{\rm 3.5PN}$ 
and $({\dot {\bf {\cal S}}}_2)_{\rm 3.5PN}$ from Eq.
(\ref{sdot35}) exactly balances a corresponding effect in 
the orbital part, so that the net ${\dot {\bf J}}$ matches the flux modulo a
total time derivative.  Thus we have established a proper energy
and angular momentum balance between the radiation flux and the evolution of
the orbit, including spin-spin effects.

\section{Conclusions}
\label{sec:discussion}

We have derived the equations of motion for binary systems of spinning
bodies from first principles, including the
effects of 
gravitational radiation reaction, and incorporating the contributions of
spin-spin coupling at 3.5PN order.  We found that the spin-spin coupling
combined with
radiation reaction leads to a small 3.5PN-order
precession of the individual spins.  
The resulting equations of motion are
instantaneous, dynamical equations, and do not rely on assumptions of energy
balance, or orbital averaging.  They may be used to study the effects of
spin-spin interactions on the inspiral of compact binaries numerically.  
We have focussed attention on effects involving products $S_1 S_2$ of the
spins; effects depending quadratically on the individual spins 
can in principle also be calculated with our approach.  This
will be the subject of future work.

\acknowledgments

This work is supported in part by the National Science Foundation under
grant number PHY 03-53180, and by the National Aeronautics
and Space Administration under grant number NNG06GI60G.  
CMW is grateful to the  
Groupe Gravitation Relativiste et Cosmologie (GR$\varepsilon$CO) of the 
Institut d'Astrophysique de Paris for its hospitality while
this work was being completed.

\appendix

\section{Extracting total time derivatives}
\label{app:timederiv}

Using the Newtonian equations of motion plus the 1PN spin-spin terms, it is
straightforward to establish a number of  identities, which may be used to
extract time derivatives from 2.5PN and 3.5PN terms in the expressions
(\ref{spindotbig}) and (\ref{dotEJ}).
For any non-negative integers $s$, $p$ and $q$, we obtain
\bea
\frac{d}{dt} \left ( \frac{v^{2s} {\dot r}^p}{r^q} \right ) &=&
\frac{v^{2s-2} {\dot r}^{p-1}}{r^{q+1}}
\left \{ pv^4 - (p+q)v^2{\dot r}^2 - 2s{\dot r}^2\frac{m}{r}
-pv^2\frac{m}{r} 
- 3p\frac{v^2}{\mu r^3} \left ( \Ss - 3 \Nsa\Nsb \right )
\right .
\nonumber
\\
&& 
\left .
\quad 
-6s \frac{\Nv}{\mu r^3} \left ( \Nv\Ss + \Vsa\Nsb + \Vsb\Nsa -5\Nv\Nsa\Nsb 
\right )
\right \}
\,,
\nonumber
\\
\frac{d}{dt} \left ( \frac{v^{2s} {\dot r}^p}{r^q} {\bf {\tilde L}}_{\rm N}
\right ) &=&
\frac{v^{2s-2} {\dot r}^{p-1}}{r^{q+1}}
\left \{
\left [  pv^4 - (p+q)v^2{\dot r}^2 - 2s{\dot r}^2\frac{m}{r}
-pv^2\frac{m}{r} 
- 3p\frac{v^2}{\mu r^3} \left ( \Ss - 3 \Nsa\Nsb \right )
\right .
\right .
\nonumber
\\
&& \left .\left .
\quad 
-6s \frac{\Nv}{\mu r^3} \left ( \Nv\Ss + \Vsa\Nsb + \Vsb\Nsa -5\Nv\Nsa\Nsb 
\right )
\right ] {\bf {\tilde L}}_{\rm N}
\right .
\nonumber \\
&&
\left .
-\frac{v^2{\dot r}}{\mu r^2} 
\left ( \NcSa \Nsb + \NcSb \Nsa \right )
\right \} \,.
\label{timederiv1}
\eea
Another set of identities, to be used only in 3.5PN terms, requires only the 
Newtonian equations of motion:
\bea
\frac{d}{dt} \left ( \frac{v^{2s} {\dot r}^p}{r^q} x^ix^j \right ) &=&
\frac{v^{2s-2} {\dot r}^{p-1}}{r^{q+1}}
\left \{  \left [pv^4 - (p+q)v^2{\dot r}^2 - 2s{\dot r}^2\frac{m}{r}
-pv^2\frac{m}{r} \right ] x^ix^j
\right .
\nonumber \\
&&
\left .
+2 v^2{\dot r} r x^{(i}v^{j)} \right \} \,,
\nonumber \\
\frac{d}{dt} \left ( \frac{v^{2s} {\dot r}^p}{r^q} v^iv^j \right ) &=&
\frac{v^{2s-2} {\dot r}^{p-1}}{r^{q+1}}
\left \{  \left [pv^4 - (p+q)v^2{\dot r}^2 - 2s{\dot r}^2\frac{m}{r}
-pv^2\frac{m}{r} \right ] v^iv^j
\right .
\nonumber \\
&&
\left .
-2m \frac{v^2{\dot r}}{r^2} x^{(i}v^{j)} \right \} \,,
\nonumber \\
\frac{d}{dt} \left ( \frac{v^{2s} {\dot r}^p}{r^q} x^iv^j \right ) &=&
\frac{v^{2s-2} {\dot r}^{p-1}}{r^{q+1}}
\left \{  \left [pv^4 - (p+q)v^2{\dot r}^2 - 2s{\dot r}^2\frac{m}{r}
-pv^2\frac{m}{r} \right ] x^iv^j
\right .
\nonumber \\
&&
\left .
+ v^2{\dot r} r \left ( v^iv^j - \frac{m}{r} n^in^j \right ) \right \} \,.
\label{timederiv2}
\eea

\end{document}